\documentclass{Interspeech2024}

\interspeechcameraready

\usepackage{amsmath}
\usepackage{graphicx}
\usepackage{comment}
\usepackage{booktabs}
\usepackage{amssymb}
\usepackage{multirow}
\usepackage{newtxtext,newtxmath}

\usepackage{hyperref}
\usepackage{cite}

\renewcommand{\Vec}[1]{\textrm{\boldmath $#1$}} % Vector

\title{SpeechBERTScore: Reference-Aware Automatic Evaluation of Speech Generation Leveraging NLP Evaluation Metrics}

\address{
  $^1$First Affiliation, CountryX\\
  $^2$Second Affiliation, CountryY \\
  $^3$Third Affiliation, CountryZ}
\email{first@university.edu, second@companyA.com, third@companyB.ai}

\name{Takaaki Saeki$^1$, Soumi Maiti$^2$, Shinnosuke Takamichi$^{1,3}$, Shinji Watanabe$^2$, Hiroshi Saruwatari$^1$}

\address{
  $^1$The University of Tokyo, Japan.\\
  $^2$Carnegie Mellon University, USA. $^3$Keio University, Japan.}
\email{saefrospace@gmail.com, smaiti@andrew.cmu.edu, shinnosuke\_takamichi@ipc.i.u-tokyo.ac.jp, shinjiw@ieee.org, hiroshi\_saruwatari@ipc.i.u-tokyo.ac.jp}


\begin{document}

\maketitle

\begin{abstract}

%While subjective assessments have been the gold standard for evaluating speech generation, objective metrics such as mel cepstral distortion (MCD) have also been used.
%Due to their cost efficiency, there is a need to establish objective metrics that are highly correlated with human subjective judgments.
While subjective assessments have been the gold standard for evaluating speech generation, there is a growing need for objective metrics that are highly correlated with human subjective judgments due to their cost efficiency.
This paper proposes reference-aware automatic evaluation methods for speech generation inspired by evaluation metrics in natural language processing.
The proposed \textit{SpeechBERTScore} computes the BERTScore for self-supervised dense speech features of the generated and reference speech, which can have different sequential lengths.
We also propose \textit{SpeechBLEU} and \textit{SpeechTokenDistance}, which are computed on speech discrete tokens.
The evaluations on synthesized speech show that our method correlates better with human subjective ratings than mel cepstral distortion and a recent mean opinion score prediction model.
Also, they are effective in noisy speech evaluation and have cross-lingual applicability.
\end{abstract}

\vspace{-1mm}
\section{Introduction}\label{sec:intro}

Subjective listening tests have been the gold standard for evaluating the quality of generated and degraded speech~\cite{black05blizzard,ITU1996P800}.
%However, due to the time consuming nature of listening tests and the inability to compare different listening test results, various objective evaluation metrics have been used.
However, various objective evaluation metrics have also been employed to reduce time and costs.
For example, objective metrics~\cite{rix2001perceptual,taal2011algorithm} such as mel cepstral distortion (MCD)~\cite{fukada92melcep} are used to compare reference and generated speech.
However, these metrics, which are based on simple acoustic features, can deviate from human subjective ratings, especially for utterances with different acoustic and prosodic characteristics but high naturalness.
Consequently, recent studies have also focused on frameworks that predict subjective ratings from input speech using neural models~\cite{lo2019mosnet,Cooper2021GeneralizationAO,mittag2021nisqa,reddy2021dnsmos,saeki2022utmos}.
However, the supervised learning based methods degrade the performance in mismatched conditions, which limits their practicality~\cite{Cooper2021GeneralizationAO}.
%However, their performance drops for data from domains other than those used in the downstream training, which limits their practicality~\cite{Cooper2021GeneralizationAO}.

In natural language processing (NLP), automatic evaluation metrics that highly correlate with human subjective judgement~\cite{zhao2019moverscore,yuan2021bartscore} have been proposed.
BLEU~\cite{papineni2002bleu} measures content agreement through $n$-gram overlaps, while BERTScore~\cite{zhang2019bertscore} assesses contextual meaning via language model semantics. 
%BLEU~\cite{papineni2002bleu} captures content agreement by counting $n$-gram overlaps between reference and generated text.
%BERTScore~\cite{zhang2019bertscore} uses semantic representations from language models to capture contextual meaning..
Recent speech self-supervised learning (SSL) models can generate semantic continuous vectors or discrete tokens from speech~\cite{baevski2020wav2vec,hsu2021hubert,chen2022wavlm}.
% Such representations allow speech tokens to be treated like text tokens, thus enabling the construction of language models on speech corpora.
Such speech representations can be treated similarly to text representations, thus enabling applications of the above NLP metrics to speech.
Our goal is to develop evaluation metrics that better match human subjective judgments using semantic speech representations.

We propose new reference-aware evaluation metrics for speech generation, which can be used when the generated and reference speech have different sequence lengths.
The proposed \textit{SpeechBERTScore} calculates the BERTScore for SSL features of the generated and reference speech, capturing their semantic congruence.
Our \textit{SpeechBLEU} and \textit{SpeechTokenDistance} calculate BLEU and character-level distance, respectively, for discrete speech tokens.
Our automatic evaluation framework using NLP metrics has the potential for future extension to include other metrics such as ROUGE~\cite{lin2004rouge} and CIDEr~\cite{vedantam2015cider}.
Table~\ref{tab:comparison-sqa} shows the difference between previous evaluation frameworks and our framework.
Unlike traditional metrics based on signal processing~\cite{fukada92melcep,rix2001perceptual}, our method uses SSL speech features for a more semantically informed evaluation.
Also, unlike previous data-driven evaluation frameworks~\cite{lo2019mosnet,saeki2022utmos}, our reference-aware approach eliminates the need for downstream training, thus lowering costs and avoiding the problem of mismatched conditions.
The evaluations demonstrate that our method gives a higher correlation with human ratings than previous automatic evaluation metrics.
%The proposed metrics will be made available as a standalone open-source toolkit and in ESPnet~\cite{watanabe2018espnet}.
Our metrics are available as a standalone toolkit\footnote{\scriptsize{\url{https://github.com/Takaaki-Saeki/DiscreteSpeechMetrics}}} and ESPnet~\cite{watanabe2018espnet}.
The contributions are as follows:
%\vspace{-1.5mm}
\begin{itemize}
    \item We propose novel automatic evaluation methods for speech generation, inspired by text generation metrics.
    \item The effectiveness of our method is confirmed through the quality evaluation of synthesized and noisy speech.
    \item Experiments using English and Chinese datasets demonstrate high cross-lingual applicability. %Surprisingly, it also correlates with subjective ratings even when using reference speech with different spoken content.
    \item Various ablation studies were conducted to investigate the effect of layer selection, token vocabulary size, etc.
\end{itemize}
%\vspace{-1.5mm}

\begin{table}[tb]
\centering
\caption{Comparison of objective speech evaluation frameworks.}
\vspace{-2mm}
\label{tab:comparison-sqa}
\scalebox{0.72}{
\begin{tabular}{l|c|c|c|c}
\toprule
Method         & \begin{tabular}[c]{@{}c@{}}Need\\ reference\end{tabular} & \begin{tabular}[c]{@{}c@{}}Need\\ labelled data\end{tabular} & \begin{tabular}[c]{@{}c@{}}Use SSL\\ pretraining\end{tabular} & \begin{tabular}[c]{@{}c@{}}Need down\\ stream training\end{tabular} \\ \midrule
MCD~\cite{fukada92melcep}, PESQ~\cite{rix2001perceptual}  & Y                                                          & N            & N                                                         & N                                                              \\
SpeechLMScore~\cite{maiti2022speechlmscore}  & N                                                          & N            & Y                                                         & N                                                              \\
MOSNet~\cite{lo2019mosnet}         & N                                                          & Y            & N                                                         & Y                                                              \\
UTMOS~\cite{saeki2022utmos} & N                                                          & Y            & Y                                                         & Y                                                              \\ \midrule
Ours           & Y                                                          & N            & Y                                                         & N                                                              \\ \bottomrule
\end{tabular}
}
\vspace{-4mm}
\end{table}

\vspace{-1mm}
\section{Method}\label{sec:method}
In this section, we describe the proposed reference-aware objective metrics.
SpeechBERTScore (\S~\ref{sec:method-bert}) is defined for SSL speech features, while SpeechBLEU (\S~\ref{sec:method-bleu}) and SpeechTokenDistance (\S~\ref{sec:method-dist}) are defined for discrete tokens.

\begin{figure}
    \centering
    \includegraphics[width=0.88\linewidth, clip]{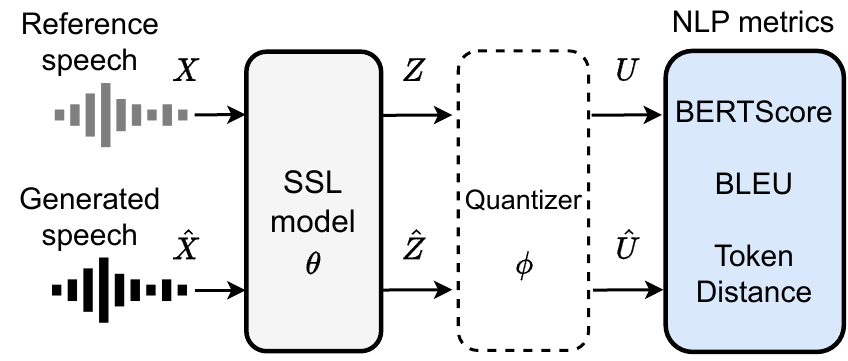}
    \caption{Proposed speech evaluation metrics. SpeechBERTScore is computed with dense SSL speech features. Quantizer is used for SpeechBLEU and SpeechTokenDistance. $Z$ and $\hat{Z}$ are SSL features. $U$ and $\hat{U}$ are speech discrete tokens.}
    \label{fig:overview}
    \vspace{-3mm}
\end{figure}

\vspace{-1mm}
\subsection{SpeechBERTScore}\label{sec:method-bert}

BERTScore~\cite{zhang2019bertscore} is a widely used automatic evaluation method for text generation.
In BERTScore, the similarity between the generated text and the reference text is calculated based on the semantic BERT~\cite{devlin2019bert} embeddings corresponding to each text token.
In this study, we propose \textit{SpeechBERTScore}, which employs BERTScore as an evaluation metric for speech generation.
\textit{SpeechBERTScore} calculates the BERTScore for SSL feature sequences from both generated and reference speech, capturing their semantic congruence.

Let $\hat{X} = (\hat{x}_{t} \in \mathbb{R} | t=1, \cdots, T_{\mathrm{gen}})$ and $X = (x_{t} \in \mathbb{R} | t=1, \cdots, T_{\mathrm{ref}})$ denote the generated and reference speech waveforms, respectively.
Here, the waveform lengths $T_{\mathrm{gen}}$ and $T_{\mathrm{ref}}$ can be different.
Let $\hat{Z} = (\hat{\Vec{z}}_{n} \in \mathbb{R}^{D} | n=1, \cdots, N_{\mathrm{gen}})$ and $Z = (\Vec{z}_{n} \in \mathbb{R}^{D} | n=1, \cdots, N_{\mathrm{ref}})$ denote the speech SSL features obtained from $\hat{X}$ and $X$, respectively, as follows.
\begin{equation}\label{eq:bertscore-enc}
\begin{split}
\hat{Z} = \operatorname{Encoder}(\hat{X}; \theta), \qquad
Z = \operatorname{Encoder}(X; \theta),
\end{split}
\end{equation}
where $\theta$ denotes model parameters of a pretrained encoder model.
$N_{\mathrm{gen}}$ and $N_{\mathrm{ref}}$ are uniquely determined by $T_{\mathrm{gen}}$ and $T_{\mathrm{ref}}$ respectively, depending on the subsampling rate of the encoder.
As in \S~\ref{sec:eval-setting-model}, we use encoder models pretrained by SSL~\cite{baevski2020wav2vec,hsu2021hubert}.

While the original BERTScore defines precision, recall and F1-score, we use the precision as we found that it performed the best in our preliminary experiment (Appendix~\ref{sec:appendix-bertscore}).
Then the SpeechBERTScore is defined on the semantic features obtained by Eq.~\eqref{eq:bertscore-enc} as:
\begin{align}
    \text{SpeechBERTScore} &= \frac{1}{N_{\mathrm{gen}}} \sum_{i=1}^{N_{\mathrm{gen}}} \max_{j} \cos(\hat{\Vec{z}}_{i}, \Vec{z}_{j})
\end{align}
where $\cos(\cdot)$ is the cosine similarity between two features.

\vspace{-1mm}
\subsection{SpeechBLEU}\label{sec:method-bleu}

BLEU~\cite{papineni2002bleu}, which computes a score based on the precision of matching $n$-grams, is a common metric for evaluating the quality of machine-translated text against human translations.
In the proposed SpeechBLEU, BLEU is calculated for discrete speech tokens to evaluate the quality of generated speech against reference speech.

Let $\hat{U} = (\hat{u}_{n} \in \mathcal{V} | n=1, \cdots, N_{\mathrm{gen}})$ and $U = (u_{n} \in \mathcal{V} | n=1, \cdots, N_{\mathrm{ref}})$ denote the discrete unit sequences obtained from $\hat{Z}$ and $Z$ (Eq.~\eqref{eq:bertscore-enc}), respectively.
Here $\mathcal{V}$ denotes the vocabulary of discrete tokens, with the vocabulary size $K$.
Using an external quantizer, the discrete units can be obtained as follows.
\begin{equation}\label{eq:quantizer}
\begin{split}
\hat{U} = \operatorname{Quantizer}(\hat{Z}; \phi), \qquad
U = \operatorname{Quantizer}(Z; \phi),
\end{split}
\end{equation}
where $\phi$ is the parameters of the quantizer.
We use a $k$-means algorithm~\cite{hartigan1979algorithm} for the quantizer.
Then the SpeechBLEU is defined on discrete tokens obtained by Eq.~\eqref{eq:quantizer} as follows:
\begin{equation}\label{eq:bleu}
\text{SpeechBLEU} = \operatorname{BLEU}(\hat{U}, U).
\end{equation}
We use the uniform weight to aggregate the BLEU scores for each $n$-gram, where the maximum $n$ is denoted as $G$.
%where $\Vec{w}$ denotes the weights for each $n$-gram.
%Let $G$ denote the maximum $n$-gram used to calculate BLEU, where $G \in \mathbb{Z}, G \geq 1$ is satisfied.
%$\Vec{w}$ is written as $\Vec{w} = (w_{g} \in \mathbb{R} | g=1, \cdots, G)$.
%Here, we use the uniform weight for each $n$-gram as $w_{g} = 1/G$.

\vspace{-1mm}
\subsection{SpeechTokenDistance}\label{sec:method-dist}

%Unlike NLP metrics that capture linguistic structure mentioned in \S~\ref{sec:method-bert} and \ref{sec:method-bleu}, primitive metrics that focus on character-level similarity have also been used.
In this study, we evaluate generated speech by calculating such token-level distances on speech discrete token sequences (referred to as \textit{SpeechTokenDistance}).
While the metrics described in \S~\ref{sec:method-bert} and \ref{sec:method-bleu} could capture long-contextual linguistic structure, SpeechTokenDistance focuses on a token-level string matching.
SpeechBERTScore and SpeechBLEU are applicable even with different order relations between reference and generated speech features, while SpeechTokenDistance assumes matching order relations.
While various token-level distance measures have been proposed, we explore the representative Levenshtein distance~\cite{levenshtein1966binary} and Jaro-Winkler distance~\cite{winkler1990string}.

%The Levenshtein distance calculates the minimum number of single-character edits (insertions, deletions, substitutions) required to change one text into another.
The Levenshtein distance calculates the minimum number of single-token edits required to change one text into another.
%This captures the superficial errors between two texts rather than semantic difference, and is often used in evaluating errors in automatic speech recognition and optical character recognition.
In the Jaro-Winkler distance, the Jaro distance~\cite{jaro1989advances} calculates similarity based on the number and order of shared tokens, while the Winkler extension~\cite{winkler1990string} gives more weight to prefixes.
The SpeechTokenDistance is computed on discrete tokens, which are obtained by Eq.~\eqref{eq:quantizer}, as follows:
\begin{equation}
\text{SpeechTokenDistance} = \operatorname{DistanceMeasure}(\hat{U}, U).
\end{equation}

\vspace{-1mm}
\section{Experimental evaluations}\label{sec:eval}

\subsection{Experimental settings}\label{sec:eval-setting}

\subsubsection{Evaluation criteria}\label{sec:eval-setting-criteria}

We evaluated the correlation of each metric and the human subjective ratings using both the linear correlation coefficient (LCC) and Spearman's rank correlation coefficient (SRCC).
Reference-aware metrics were computed by using both generated and reference speech, while reference-free metrics were computed only using generated speech.
Both utterance-level and system-level metrics were used to evaluate synthesized speech, while only uttreance-level metrics were evaluated for noisy speech.
%To evaluate synthesized speech, both utterance-level and system-level metrics were used, where system-level metrics are calculated by averaging utterance-level metrics for each system.
%For noisy speech, only utterance-level metrics were used.
Note that a lower MCD indicates better results, while a higher SpeechBERTScore is preferable.
Therefore, we used the absolute LCC and SRCC values in our evaluation.

\subsubsection{Dataset}\label{sec:eval-setting-dataset}

We used three types of evaluation datasets, where the sampling rate was set to 16~kHz.
We used the SOMOS dataset~\cite{maniati2022somos} for the evaluation of English synthesized speech.
It contains LJSpeech~\cite{ljspeech17} voices synthesized by 200 different text-to-speech (TTS) acoustic models and the LPCNet~\cite{valin2019lpcnet} vocoder, along with the corresponding ratings.
Since our evaluation requires reference speech, we used only the LJSpeech domain and employed the SOMOS-clean subset, which applies quality filtering to the ratings.
Consequently, the dataset included 1000 synthetic utterances, their corresponding ratings and natural speech utterances.

To investigate the cross-lingual applicability of our method, we evaluated Chinese synthesized speech.
We used the Blizzard Challenge 2019 (BC2019)~\cite{wu2019blizzard} subset of the BVCC dataset~\cite{cooper2021bvcc}, which provides 1300 synthesised speech samples, along with their corresponding ratings and natural speech utterances.

We also used the \texttt{NISQA\_VAL\_SIM} subset of the NISQA Corpus~\cite{mittag2021nisqa} for the evaluation of noisy speech.
It has 2500 noisy speech utterances generated with various distortions, along with their corresponding ratings and clean speech utterances.

\vspace{-1mm}
\subsubsection{Self-supervised pretrained models}\label{sec:eval-setting-model}

In the evaluation of SpeechBERTScore in \S~\ref{sec:method-bert}, we explored multiple SSL models including Wav2vec~2.0~\cite{baevski2020wav2vec}, HuBERT~\cite{hsu2021hubert}, WavLM~\cite{chen2022wavlm}, and Encodec~\cite{defossez2022high}.
We used models available in fairseq~\cite{ott2019fairseq}: Wav2vec~2.0 Base (\texttt{wav2vec2-base}), Wav2vec~2.0 Large (\texttt{wav2vec2-large}), HuBERT Base (\texttt{hubert-base}) and HuBERT Large (\texttt{hubert-large}).
We also used WavLM models available in the official repository\footnote{\scriptsize{\url{https://github.com/microsoft/unilm/tree/master/wavlm}}}: WavLM Base (\texttt{wavlm-base}), WavLM Base+ (\texttt{wavlm-base+}) and WavLM Large (\texttt{wavlm-large}).
For Encodec (\texttt{encodec})\footnote{\scriptsize{\url{https://github.com/facebookresearch/encodec}}}, continuous features before the residual vector quantization layers were employed.
In the evaluations except for \S~\ref{sec:eval-layer}, we report results with the best-performing layer.

To evaluate Chinese synthetic speech described in \S~\ref{sec:eval-setting-dataset}, we used published models\footnote{\scriptsize{\url{https://github.com/TencentGameMate/chinese_speech_pretrain}}}: Wav2vec 2.0 Base (\texttt{wav2vec2-base-cmn}) and Wav2vec 2.0 Large (\texttt{wav2vec2-large-cmn}), HuBERT Base (\texttt{hubert-base-cmn}) and HuBERT Large (\texttt{hubert-large-cmn}), as well as multilingual XLSR models~\cite{conneau2020unsupervised} trained on 53 (\texttt{xlsr-53}) and 128 languages (\texttt{xlsr-128}) available in fairseq.

For the evaluation of SpeechBLEU and SpeechTokenDistance, we transformed the \texttt{hubert-base} features into discrete tokens using a $k$-means model trained on LibriSpeech 960h~\cite{panayotov15librispeech}, comparing different vocabulary sizes $K$.
In the evaluations except for \S~\ref{sec:eval-layer}, we report results with the best-performing layer.

\begin{table}[tb]
\centering
\caption{Main results on synthetic speech (SOMOS).}
\vspace{-2mm}
\label{tab:eval-main-synth}
\scalebox{0.78}{
\begin{tabular}{lcccc}
\toprule
\multicolumn{1}{l|}{\multirow{2}{*}{}}                  & \multicolumn{2}{c|}{Utterance-level}                 & \multicolumn{2}{c}{System-level} \\
\multicolumn{1}{l|}{}                                   & LCC            & \multicolumn{1}{c|}{SRCC}           & LCC             & SRCC           \\ \midrule
\multicolumn{5}{l}{\textit{Traditional reference-aware metrics}}                                                                                  \\ \midrule
\multicolumn{1}{l|}{MCD~\cite{fukada92melcep}}                                & 0.356          & \multicolumn{1}{c|}{0.330}          & 0.541           & 0.518          \\
\multicolumn{1}{l|}{Log F0 RMSE}                        & 0.050          & \multicolumn{1}{c|}{0.057}          & 0.116           & 0.123          \\ \midrule
\multicolumn{5}{l}{\textit{Reference-free metrics, Unsupervised}}                                                                                      \\ \midrule
\multicolumn{1}{l|}{SpeechLMScore~\cite{maiti2022speechlmscore}}                      & 0.164          & \multicolumn{1}{c|}{0.127}          & 0.268           & 0.246          \\ \midrule
\multicolumn{5}{l}{\textit{Reference-free metrics, Supervised}}                                                                                        \\ \midrule
\multicolumn{1}{l|}{UTMOS~\cite{saeki2022utmos}}                              & 0.363          & \multicolumn{1}{c|}{0.340}          & 0.537           & 0.575          \\ \midrule
\multicolumn{5}{l}{\textit{Proposed (Reference-aware metrics, Unsupervised)}}                                                                     \\ \midrule
\multicolumn{1}{l|}{SpeechBERTScore}                    & \textbf{0.581} & \multicolumn{1}{c|}{\textbf{0.563}} & \textbf{0.781}  & \textbf{0.760} \\
\multicolumn{1}{l|}{SpeechBLEU ($G=2$)}                         & 0.427          & \multicolumn{1}{c|}{0.423}          & 0.680           & 0.659          \\
\multicolumn{1}{l|}{SpeechTokenDistance (Levenshtein)}  & 0.247          & \multicolumn{1}{c|}{0.210}          & 0.414           & 0.362          \\
\multicolumn{1}{l|}{SpeechTokenDistance (Jaro-Winkler)} & 0.407          & \multicolumn{1}{c|}{0.427}          & 0.663           & 0.681          \\ \bottomrule
\end{tabular}
}
\vspace{-4mm}
\end{table}

\subsubsection{Baselines}\label{sec:eval-setting-baseline}
To evaluate the synthesized speech, we used MCD~\cite{fukada92melcep} and log F0 root mean squared error (RMSE), common reference-aware metrics in speech synthesis, where we used evaluation scripts in ESPnet2-TTS~\cite{watanabe2018espnet,hayashi2021espnet2}.
%For MCD and Log F0 RMSE, we applied dynamic time-warping (DTW)~\cite{salvador2007toward} to match the length between reference and generated speech.
We used SpeechLMScore~\cite{maiti2022speechlmscore} as a reference-free unsupervised method, using a published pre-trained model\footnote{\scriptsize{\url{https://github.com/soumimaiti/speechlmscore_tool}}} trained on LibriSpeech 960h~\cite{panayotov15librispeech}.
We used the default model (\texttt{50\_3} setting in the paper) with a token vocabulary of 50 using the 3rd layer features.
As a reference-free supervised method, we used a UTMOS~\cite{saeki2022utmos} strong learner, trained on the BVCC dataset, which was one of the top performing models in the VoiceMOS Challenge 2022~\cite{huang21voicemos}.
Note that this is an out-of-domain prediction setting, which resulted in a lower correlation than the in-domain prediction in the SOMOS paper~\cite{maniati2022somos}.

For the evaluation of noisy speech, we used popular reference-aware metrics such as Perceptual evaluation of speech quality (PESQ)~\cite{rix2001perceptual}, a short-time objective intelligibility measure (STOI)~\cite{taal2011algorithm}, extended STOI (ESTOI)~\cite{jensen2016algorithm} and signal-to-distortion ratio (SDR).
We also used the SpeechLMScore~\cite{maiti2022speechlmscore} model, which is used to evaluate synthesized speech.
We used a pretrained DNSMOS model~\cite{reddy2021dnsmos} as a reference-free supervised model, which includes three metrics: signal quality (SIG), background quality (BAK) and overall quality (OVRL).

\vspace{-1mm}
\subsection{Main results}\label{sec:eval-main}

\begin{table}[tb]
\centering
\caption{Main results on noisy speech (NISQA Corpus). \textit{SpTokDis} denotes SpeechTokenDistance defined in \S~\ref{sec:method-dist}. \textit{Leven.} and \textit{J.-W.} denote Levenshtein and Jaro-Winkler, respectively.}
\vspace{-2mm}
\label{tab:eval-main-noisy}
\scalebox{0.75}{
\begin{tabular}{lcccccc}
\toprule
\multicolumn{1}{l|}{\multirow{2}{*}{}}           & \multicolumn{2}{c|}{Aligned ref.}                    & \multicolumn{2}{c|}{0.99x ref.}                      & \multicolumn{2}{c}{1.01x ref.}  \\
\multicolumn{1}{l|}{}                            & LCC            & \multicolumn{1}{c|}{SRCC}           & LCC            & \multicolumn{1}{c|}{SRCC}           & LCC            & SRCC           \\ \midrule
\multicolumn{7}{l}{\textit{Traditional reference-aware metrics}}                                                                                                                                 \\ \midrule
\multicolumn{1}{l|}{PESQ}                        & \textbf{0.841} & \multicolumn{1}{c|}{0.840}          & 0.678          & \multicolumn{1}{c|}{0.667}          & 0.686          & 0.673          \\
\multicolumn{1}{l|}{STOI}                        & 0.741          & \multicolumn{1}{c|}{0.825}          & 0.268          & \multicolumn{1}{c|}{0.251}          & 0.348          & 0.337          \\
\multicolumn{1}{l|}{ESTOI}                       & 0.764          & \multicolumn{1}{c|}{0.826}          & 0.369          & \multicolumn{1}{c|}{0.339}          & 0.506          & 0.489          \\
\multicolumn{1}{l|}{SDR}                         & 0.346          & \multicolumn{1}{c|}{0.741}          & 0.126          & \multicolumn{1}{c|}{0.112}          & 0.289          & 0.264          \\ \midrule
\multicolumn{7}{l}{\textit{Reference-free metrics, Unsupervised}}                                                                                                                                \\ \midrule
\multicolumn{1}{l|}{SpeechLMScore}               & 0.583          & \multicolumn{1}{c|}{0.583}          & 0.583          & \multicolumn{1}{c|}{0.583}          & 0.583          & 0.583          \\ \midrule
\multicolumn{7}{l}{\textit{Reference-free metrics, Supervised}}                                                                                                                                  \\ \midrule
\multicolumn{1}{l|}{DNSMOS (BAK)}                & 0.542          & \multicolumn{1}{c|}{0.567}          & 0.542          & \multicolumn{1}{c|}{0.567}          & 0.542          & 0.567          \\
\multicolumn{1}{l|}{DNSMOS (SIG)}                & 0.595          & \multicolumn{1}{c|}{0.642}          & 0.595          & \multicolumn{1}{c|}{0.642}          & 0.595          & 0.642          \\
\multicolumn{1}{l|}{DNSMOS (OVRL)}               & 0.674          & \multicolumn{1}{c|}{0.697}          & 0.674          & \multicolumn{1}{c|}{0.697}          & 0.674          & 0.697          \\ \midrule
\multicolumn{7}{l}{\textit{Proposed (Reference-aware metrics, Unsupervised)}}                                                                                                                    \\ \midrule
\multicolumn{1}{l|}{SpeechBERTScore}             & 0.824          & \multicolumn{1}{c|}{\textbf{0.868}} & 0.738          & \multicolumn{1}{c|}{\textbf{0.793}} & 0.747          & \textbf{0.801} \\
\multicolumn{1}{l|}{SpeechBLEU}                  & 0.821          & \multicolumn{1}{c|}{0.827}          & \textbf{0.747} & \multicolumn{1}{c|}{0.738}          & \textbf{0.754} & 0.744          \\
\multicolumn{1}{l|}{SpeechTokDis (Leven.)}  & 0.762          & \multicolumn{1}{c|}{0.800}          & 0.633          & \multicolumn{1}{c|}{0.637}          & 0.637          & 0.644          \\
\multicolumn{1}{l|}{SpeechTokDis (J.-W.)} & 0.778          & \multicolumn{1}{c|}{0.777}          & 0.707          & \multicolumn{1}{c|}{0.729}          & 0.723          & 0.732          \\ \bottomrule
\end{tabular}
}
\vspace{-4mm}
\end{table}

We first conducted evaluations on English synthetic speech as in \S~\ref{sec:eval-setting-dataset}.
For SpeechBERTScore, we used the \texttt{wavlm-large} model.
We also used SpeechBLEU with no token repetition, $K=200$ and the $G=2$ setting, while we used SpeechTokenDistance with token repetition and $K=200$.
Table~\ref{tab:eval-main-synth} lists the results.
Our metrics, SpeechBERTScore, SpeechBLEU, and SpeechTokenDistance (Jaro-Winkler), outperformed traditional reference-aware metrics in all criteria.
They also showed higher correlations than the unsupervised SpeechLMScore and the supervised UTMOS.
There was no overlap in the 95\% confidence intervals between the proposed SpeechBERTScore and previous metrics for both the utterance-level and system-level scores.
These results highlight the effectiveness of our metrics in better aligning with human subjective ratings, with SpeechBERTScore exhibiting the highest correlation.

For noisy speech evaluations described in \S~\ref{sec:eval-setting-dataset}, we used the same configurations in the proposed methods as for the synthetic speech evaluations.
Note that the original reference and noisy speech utterances are time aligned.
Results of \textit{Aligned ref.} in Table~\ref{tab:eval-main-noisy} reveal that while PESQ had the highest LCC, our SpeechBERTScore achieved the highest SRCC.
SpeechBERTScore and SpeechBLEU outperformed other methods\footnote{The low LCC of SDR is presumably due to its wider range of values.} in both LCC and SRCC, except for PESQ.
While signal-processing-based methods exhibited high performance in \textit{Aligned ref.}, our method can also be used under conditions where the reference and noisy speech utterances are not time-aligned.
To simulate cases where they are not time-aligned, we conducted evaluations using reference speech utterance that was time-stretched to 0.99 times (\textit{0.99x ref.}) or 1.01 times (\textit{1.01x ref.}). As a result in Table~\ref{tab:eval-main-noisy}, our SpeechBERTScore and SpeechBLEU were found to be more robust to unaligned references than PESQ.

\vspace{-1mm}
\subsection{Ablation study of speech-token-based metrics}\label{sec:eval-token}
\vspace{-1mm}

\begin{table}[tb]
\centering
\caption{Investigation of token repetition and vocabulary for speech-token-based metrics.}
\vspace{-2mm}
\label{tab:eval-token-km}
\scalebox{0.84}{
\begin{tabular}{l|l|cc|cc}
\toprule
\multirow{2}{*}{Repetition}                   & \multirow{2}{*}{Vocab.} & \multicolumn{2}{c|}{\begin{tabular}[c]{@{}c@{}}Utterance-level\\ SpeechBLEU\\ ($G=2$)\end{tabular}} & \multicolumn{2}{c}{\begin{tabular}[c]{@{}c@{}}Utterance-level\\ SpeechTokenDistance\\ (Jaro-Winkler)\end{tabular}} \\
                                    &                               & LCC                                                & SRCC                                               & LCC                                                      & SRCC                                                    \\ \midrule
\multirow{3}{*}{w/ rep}    & $K=50$                          & 0.341                                              & 0.325                                              & 0.284                                                    & 0.275                                                   \\
                                    & $K=100$                         & 0.364                                              & 0.346                                              & 0.354                                                    & 0.356                                                   \\
                                    & $K=200$                         & 0.407                                              & 0.396                                              & \textbf{0.407}                                           & \textbf{0.427}                                          \\ \midrule
\multirow{3}{*}{w/o rep} & $K=50$                          & 0.357                                              & 0.342                                              & 0.202                                                    & 0.202                                                   \\
                                    & $K=100$                         & 0.386                                              & 0.369                                              & 0.304                                                    & 0.329                                                   \\
                                    & $K=200$                         & \textbf{0.427}                                     & \textbf{0.423}                                     & 0.370                                                    & 0.379                                                   \\ \bottomrule
\end{tabular}
}
\vspace{-1mm}
\end{table}

We conducted an ablation study on the proposed speech-token-based metrics described in \S~\ref{sec:method-bleu} and \ref{sec:method-dist}.
While previous studies~\cite{lakhotia2021generative,maiti2022speechlmscore} have removed the repetition of discrete tokens to reduce the redundancy and overall sequence length, it can ignore the duration information of speech tokens.
We thus compared cases with (\texttt{w/ rep}) and without (\texttt{w/o rep}) speech token repetition.
Following previous work~\cite{maiti2022speechlmscore}, we varied the token vocabulary size $K$ to 50, 100, and 200.
We used $G=2$ defined in \S~\ref{sec:method-bleu}, as it gave the best results in \S~\ref{sec:appendix-ngram}.

Results in Table~\ref{tab:eval-token-km} indicate higher correlations for both metrics with increased vocabulary size, highlighting the effectiveness of using tokens with richer speech information.
For SpeechBLEU, \texttt{w/o rep} showed better performance, while for SpeechTokenDistance, \texttt{w/ rep} was more effective.
Token duration may be important for calculating character-level similarity in SpeechTokenDistance, whereas for SpeechBLEU, removing token repetition may better capture content similarity.

\vspace{-1mm}
\subsection{Layer-wise analysis}\label{sec:eval-layer}

\begin{figure}
    \centering
    \includegraphics[width=1.0\linewidth, clip]{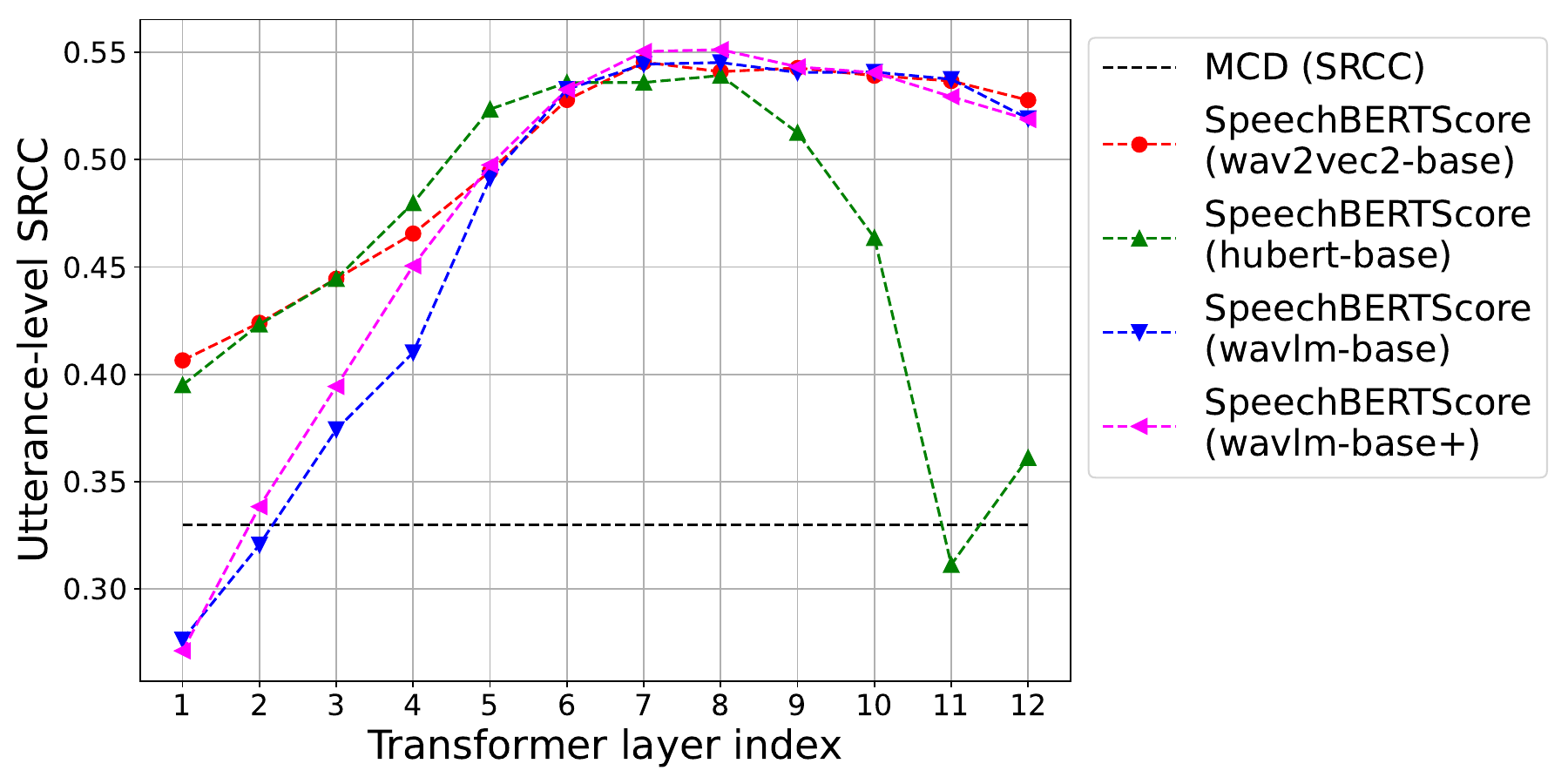}
    \vspace{-4mm}
    \caption{Analysis of layers in SSL models.}
    \label{fig:eval-layer}
    \vspace{-4mm}
\end{figure}

We investigated the effect of Transformer layer features from SSL models on SpeechBERTScore described in \S~\ref{sec:method-bert}.
Fig.~\ref{fig:eval-layer} plots the utterance-level SRCCs using features from different Transformer layer indices, where a lower index number corresponds to layers closer to the input.
As a result, using features from layers 1 or 2 resulted in lower correlations.
In contrast, layer indices of 3 or higher generally led to larger utterance-level SRCCs for SpeechBERTScore compared to MCD, suggesting that higher layers with more semantic information~\cite{pasad2021layer} are beneficial for our metrics based on NLP evaluation metrics.
Notably, the SSL models except for \texttt{hubert-base} had the beneficial property of being highly robust to layer selection, indicating that layers beyond the 7th can be selected randomly.
% In addition, there were different trends between SSL models; \texttt{wav2vec2-base} and \texttt{wavlm-base} were more robust in layer selection compared to \texttt{hubert-base}.

\vspace{-1mm}
\subsection{Model-wise analysis}\label{sec:eval-model}

\begin{table}[tb]
\centering
\caption{Model-wise analysis for SOMOS (English).}
\vspace{-2mm}
\label{tab:eval-model-en}
\scalebox{0.90}{
\begin{tabular}{lcccc}
\toprule
\multicolumn{1}{l|}{\multirow{3}{*}{}} & \multicolumn{4}{c}{SOMOS (English)}                                                     \\
\multicolumn{1}{l|}{}                  & \multicolumn{2}{c|}{Utterance-level}                 & \multicolumn{2}{c}{System-level} \\
\multicolumn{1}{l|}{}                  & LCC            & \multicolumn{1}{c|}{SRCC}           & LCC             & SRCC           \\ \midrule
\multicolumn{5}{l}{\textit{Traditional reference-aware metrics}}                                                                 \\ \midrule
\multicolumn{1}{l|}{MCD~\cite{fukada92melcep}}               & 0.356          & \multicolumn{1}{c|}{0.330}          & 0.541           & 0.518          \\ \midrule
\multicolumn{5}{l}{\textit{Proposed reference-aware metrics (SpeechBERTScore)}}                                                  \\ \midrule
\multicolumn{1}{l|}{\texttt{encodec}}           & 0.087          & \multicolumn{1}{c|}{0.074}          & 0.158           & 0.144          \\
\multicolumn{1}{l|}{\texttt{wav2vec2-base}}  & 0.560          & \multicolumn{1}{c|}{0.539}          & 0.776           & 0.745          \\
\multicolumn{1}{l|}{\texttt{wav2vec2-large}} & 0.566          & \multicolumn{1}{c|}{0.547}          & 0.770           & 0.744          \\
\multicolumn{1}{l|}{\texttt{hubert-base}}       & 0.564          & \multicolumn{1}{c|}{0.545}          & 0.775           & 0.740          \\
\multicolumn{1}{l|}{\texttt{hubert-large}}      & 0.563          & \multicolumn{1}{c|}{0.548}          & 0.766           & 0.730          \\
\multicolumn{1}{l|}{\texttt{wavlm-base}}        & 0.559          & \multicolumn{1}{c|}{0.545}          & 0.769           & 0.739          \\
\multicolumn{1}{l|}{\texttt{wavlm-base+}}       & 0.566          & \multicolumn{1}{c|}{0.551}          & 0.767           & 0.741          \\
\multicolumn{1}{l|}{\texttt{wavlm-large}}       & \textbf{0.581} & \multicolumn{1}{c|}{\textbf{0.563}} & \textbf{0.781}  & \textbf{0.760} \\ \bottomrule
\end{tabular}
}
\vspace{-1mm}
\end{table}

In evaluating English synthetic speech, we compared different SSL models mentioned in \S~\ref{sec:eval-setting-model}, with results in Table~\ref{tab:eval-model-en}.
First, \texttt{encodec} underperformed in all metrics, indicating the importance of semantic over acoustic information for SpeechBERTScore. 
Among other models, larger models generally outperformed the base models; in particular, \texttt{wavlm-large} excelled in all metrics.
WavLM's improved learning for a wider range of speech tasks due to its extended pretext task highlights its effectiveness for SpeechBERTScore.

\begin{table}[tb]
\centering
\caption{Model-wise analysis for BC2019 (Chinese).}
\vspace{-2mm}
\label{tab:eval-model-cmn}
\scalebox{0.88}{
\begin{tabular}{lcccc}
\toprule
\multicolumn{1}{l|}{\multirow{3}{*}{}}         & \multicolumn{4}{c}{BC2019 (Chinese)}                                                    \\
\multicolumn{1}{l|}{}                          & \multicolumn{2}{c|}{Utterance-level}                 & \multicolumn{2}{c}{System-level} \\
\multicolumn{1}{l|}{}                          & LCC            & \multicolumn{1}{c|}{SRCC}           & LCC             & SRCC           \\ \midrule
\multicolumn{5}{l}{\textit{Traditional reference-aware metrics}}                                                                         \\ \midrule
\multicolumn{1}{l|}{MCD~\cite{fukada92melcep}}                       & 0.156          & \multicolumn{1}{c|}{0.300}          & 0.153           & 0.362          \\ \midrule
\multicolumn{5}{l}{\textit{Proposed reference-aware metrics (SpeechBERTScore)}}                                                          \\ \midrule
\multicolumn{1}{l|}{\texttt{wav2vec2-large}}         & 0.746          & \multicolumn{1}{c|}{0.644}          & 0.834           & 0.725          \\
\multicolumn{1}{l|}{\texttt{hubert-large}}              & 0.735          & \multicolumn{1}{c|}{0.682}          & 0.867           & 0.787          \\
\multicolumn{1}{l|}{\texttt{wavlm-large}}               & 0.748          & \multicolumn{1}{c|}{0.654}          & 0.849           & 0.755          \\ \midrule
\multicolumn{1}{l|}{\texttt{wav2vec2-large-cmn}} & 0.753          & \multicolumn{1}{c|}{0.684}          & 0.856           & 0.785          \\
\multicolumn{1}{l|}{\texttt{hubert-large-cmn}}      & \textbf{0.781} & \multicolumn{1}{c|}{0.701} & 0.879           & 0.819          \\
\multicolumn{1}{l|}{\texttt{xlsr-53}}                   & 0.750          & \multicolumn{1}{c|}{\textbf{0.706}}          & \textbf{0.904}  & 0.874          \\
\multicolumn{1}{l|}{\texttt{xlsr-128}}                  & 0.742          & \multicolumn{1}{c|}{0.679}          & 0.884           & \textbf{0.889} \\ \bottomrule
\end{tabular}
}
\vspace{-3mm}
\end{table}

We investigated the cross-lingual applicability of the proposed SpeechBERTScore using the Chinese BC2019 dataset mentioned in \S~\ref{sec:eval-setting-dataset}.
This evaluation assessed the robustness of our method using an English-only SSL model for the evaluation of Chinese synthetic speech.
The results in Table~\ref{tab:eval-model-cmn} reveal that models trained on datasets including Chinese showed better performance than those trained on English speech corpora only.
More importantly, even models trained only on English corpora outperformed MCD in all metrics, confirming the cross-lingual applicability.
This highlights the utility of our metrics for low-resource languages that lack their own SSL models.

\vspace{-1mm}
\section{Conclusions}\label{sec:concl}
\vspace{-1mm}
In this paper, we proposed speech evaluation metrics based on objective text generation metrics, including SpeechBERTScore, SpeechBLEU and SpeechTokenDistance.
Experimental evaluations showed that SpeechBERTScore correlates better with human subjective ratings than traditional reference-aware metrics and previous MOS prediction models.
Our evaluations also suggested the cross-lingual applicability, indicating high practical potential.
Future work includes exploring a wider range of text generation metrics, such as MoverScore~\cite{zhao2019moverscore}.

\noindent \textbf{Acknowledgements:}
Part of this work was supported by JSPS KAKENHI Grant Number 23H03418, 23K18474, 22H03639, 21H05054, and 22KJ0838, Moonshot R\&D Grant Number JPMJPS2011, and JST FOREST JPMJFR226V.

\bibliographystyle{IEEEtran}
\bibliography{mybib}

\appendix

\section{Other analysis of SpeechBERTScore}\label{sec:appendix-bertscore}

In \S~\ref{sec:method-bert}, we present SpeechBERTScore, which calculates the precision of BERTScore on SSL features of generated and reference speech.
The precision, recall and F1-score can be defined as follows.
\begin{align}
    \text{Precision} &= \frac{1}{N_{\mathrm{gen}}} \sum_{i=1}^{N_{\mathrm{gen}}} \max_{j} \cos(\hat{\Vec{z}}_{i}, \Vec{z}_{j}) \label{eq:spbertscore-precision} \\
    \text{Recall} &= \frac{1}{N_{\mathrm{ref}}} \sum_{j=1}^{N_{\mathrm{ref}}} \max_{i} \cos(\hat{\Vec{z}}_{i}, \Vec{z}_{j})\label{eq:spbertscore-recall} \\
    \text{F1-score} &= 2 \cdot \frac{\text{Precision} \cdot \text{Recall}}{\text{Precision} + \text{Recall}}. \label{eq:spbertscore-f1}
\end{align}
We investigate the performance of each of Eq.~\eqref{eq:spbertscore-precision}--\eqref{eq:spbertscore-f1}.

Since rare words have been shown to contribute more significantly to sentence similarity than common words~\cite{banerjee2005meteor}, the original BERTScore employs importance weighting using inverse document frequency (idf).
In this section, we also investigate importance weighting focusing on either common or rare discrete tokens.
The document frequency (df) of a unit is the count of sentences that contain the unit $u$, which is given by $\text{DF}(u) = |\{d \in D : u \in d\}|$.
Here, $D$ represents the set of all utterances, which can be a large speech corpus.
Then the idf can be formulated as $\text{idf}(u) = \log \left( \frac{N}{\text{df}(u) + 1} \right)$, where $1$ is added to avoid the zero division.
Let $w_{\mathrm{gen},i}$ denote the df/idf of $i$-th generated discrete units, where $w_{\mathrm{gen},i}=\{df(\hat{u}_{i}), idf(\hat{u}_{i})\}$ is satisfied.
Then the precision with the importance weighting can be written as
\begin{align}
    \text{Precision} &= \frac{\sum_{i=1}^{N_{\mathrm{gen}}} w_{\mathrm{gen},i} \cdot \max_{j} \cos(\hat{\Vec{z}}_{i}, \Vec{z}_{j})}{\sum_{i=1}^{N_{\mathrm{gen}}} w_{\mathrm{gen},i}}.
\end{align}

We investigated the effect of different BERTScore calculation methods (Eq. (6)--(8)) and the effect of importance weighting.
Table~\ref{tab:eval-bertscore-importance} lists the results.
We found that performance did not improve with df/idf weighting.
Furthermore, when comparing precision, recall and F1-score, precision showed the highest correlation across all metrics.

\begin{table}[tb]
\centering
\caption{Analysis of different SpeechBERTScore calculations and importance weighting.}
\label{tab:eval-bertscore-importance}
\begin{tabular}{lcccc}
\toprule
\multicolumn{1}{l|}{\multirow{2}{*}{}} & \multicolumn{2}{c|}{Utterance-level}                 & \multicolumn{2}{c}{System-level} \\
\multicolumn{1}{l|}{}                  & LCC            & \multicolumn{1}{c|}{SRCC}           & LCC             & SRCC           \\ \midrule
\multicolumn{5}{l}{\textit{W/o df/idf weight}}                                                                                   \\ \midrule
\multicolumn{1}{l|}{Precision}         & \textbf{0.564} & \multicolumn{1}{c|}{\textbf{0.545}} & \textbf{0.775}  & \textbf{0.740} \\
\multicolumn{1}{l|}{Recall}            & 0.515          & \multicolumn{1}{c|}{0.495}          & 0.718           & 0.670          \\
\multicolumn{1}{l|}{F1-score}          & 0.553          & \multicolumn{1}{c|}{0.530}          & 0.758           & 0.716          \\ \midrule
\multicolumn{5}{l}{\textit{df weight}}                                                                                           \\ \midrule
\multicolumn{1}{l|}{Precision}         & 0.562          & \multicolumn{1}{c|}{0.544}          & 0.772           & 0.737          \\
\multicolumn{1}{l|}{Recall}            & 0.515          & \multicolumn{1}{c|}{0.493}          & 0.718           & 0.669          \\
\multicolumn{1}{l|}{F1-score}          & 0.551          & \multicolumn{1}{c|}{0.528}          & 0.756           & 0.714          \\ \midrule
\multicolumn{5}{l}{\textit{idf weight}}                                                                                          \\ \midrule
\multicolumn{1}{l|}{Precision}         & 0.553          & \multicolumn{1}{c|}{0.539}          & 0.772           & 0.735          \\
\multicolumn{1}{l|}{Recall}            & 0.501          & \multicolumn{1}{c|}{0.485}          & 0.702           & 0.670          \\
\multicolumn{1}{l|}{F1-score}          & 0.548          & \multicolumn{1}{c|}{0.527}          & 0.758           & 0.723          \\ \bottomrule
\end{tabular}
\end{table}

\section{N-gram analysis of SpeechBLEU}\label{sec:appendix-ngram}

We analyzed the effect of $n$-gram configuration (i.e., $G$ for Eq.~\eqref{eq:bleu}) on SpeechBLEU.
The results, presented in Table~\ref{tab:eval-token-ngram}, show that $G=2$ performs best in terms of both LCC and SRCC.
Performance diminishes with higher $G$ values like 4 or 6.
This decrease in performance for larger $n$-grams is likely due to the rarity of matches between generated and reference speech $n$-grams, which reduces the reliability of the BLEU score.

\begin{table}[tb]
\centering
\caption{Investigation of $n$-grams for SpeechBLEU.}
\label{tab:eval-token-ngram}
\begin{tabular}{c|cc|cc}
\toprule
\multicolumn{1}{l|}{\multirow{2}{*}{$G$}} & \multicolumn{2}{c|}{Utterance-level} & \multicolumn{2}{c}{System-level} \\
\multicolumn{1}{l|}{}                        & LCC               & SRCC             & LCC             & SRCC           \\ \midrule
1                                            & 0.404             & 0.406            & 0.649           & 0.638          \\
2                                            & \textbf{0.427}    & \textbf{0.423}   & \textbf{0.680}  & \textbf{0.659} \\
4                                            & 0.407             & 0.403            & 0.674           & 0.657          \\
6                                            & 0.361             & 0.360            & 0.623           & 0.610          \\ \bottomrule
\end{tabular}
\end{table}

\section{Subword tokenization for token-based metrics}\label{sec:appendix-subword}

\begin{table}[tb]
\centering
\caption{Analysis of subword tokenization.}
\label{tab:eval-subword}
\scalebox{0.9}{
\begin{tabular}{lcccc}
\toprule
\multicolumn{1}{l|}{\multirow{2}{*}{}} & \multicolumn{2}{c|}{\begin{tabular}[c]{@{}c@{}}Utterance-level\\ SpeechBLEU\\ ($G=2$, w/o rep)\end{tabular}} & \multicolumn{2}{c}{\begin{tabular}[c]{@{}c@{}}Utterance-level\\ SpeechTokenDistance\\ (Jaro-Winkler, w/ rep)\end{tabular}} \\
\multicolumn{1}{l|}{}                  & LCC                                          & \multicolumn{1}{c|}{SRCC}                                         & LCC                                                          & SRCC                                                        \\ \midrule
\multicolumn{1}{l|}{km200}       & \textbf{0.427}                               & \multicolumn{1}{c|}{\textbf{0.423}}                               & \textbf{0.407}                                               & \textbf{0.427}                                              \\ \midrule
\multicolumn{5}{l}{\textit{BPE}}                                                                                                                                                                                                                                                       \\ \midrule
\multicolumn{1}{l|}{Vocab: 256}        & 0.390                                        & \multicolumn{1}{c|}{0.399}                                        & 0.393                                                        & 0.406                                                       \\
\multicolumn{1}{l|}{Vocab: 512}        & 0.340                                        & \multicolumn{1}{c|}{0.353}                                        & 0.364                                                        & 0.351                                                       \\
\multicolumn{1}{l|}{Vocab: 1024}       & 0.315                                        & \multicolumn{1}{c|}{0.333}                                        & 0.317                                                        & 0.309                                                       \\
\multicolumn{1}{l|}{Vocab: 4096}       & 0.245                                        & \multicolumn{1}{c|}{0.259}                                        & 0.236                                                        & 0.215                                                       \\ \midrule
\multicolumn{5}{l}{\textit{Unigram}}                                                                                                                                                                                                                                                   \\ \midrule
\multicolumn{1}{l|}{Vocab: 256}        & 0.374                                        & \multicolumn{1}{c|}{0.375}                                        & 0.371                                                        & 0.398                                                       \\
\multicolumn{1}{l|}{Vocab: 512}        & 0.342                                        & \multicolumn{1}{c|}{0.347}                                        & 0.365                                                        & 0.357                                                       \\
\multicolumn{1}{l|}{Vocab: 1024}       & 0.291                                        & \multicolumn{1}{c|}{0.296}                                        & 0.303                                                        & 0.296                                                       \\
\multicolumn{1}{l|}{Vocab: 4096}       & 0.235                                        & \multicolumn{1}{c|}{0.254}                                        & 0.231                                                        & 0.231                                                       \\ \bottomrule
\end{tabular}
}
\end{table}
 
In natural language processing, subword tokenization, which treats frequent combinations of symbols as a single token, has long been used.
A recent study~\cite{shen2023acoustic} has applied byte pair encoding (BPE) tokenization to speech tokens.
While we do not use subword tokenization in the evaluations described in \S~\ref{sec:eval}, in this section, we investigate subword tokenization based on SentencePiece~\cite{kudo2018sentencepiece} for our objective measures using speech discrete tokens.
Let $\hat{S} = (\hat{s}_{n} \in \mathbb{R} | m=1, \cdots, M_{\mathrm{gen}})$ and $S = (s_{n} \in \mathbb{R} | n=1, \cdots, M_{\mathrm{ref}})$ denote the subword sequences obtained from $\hat{U}$ and $U$.
\begin{equation}
\begin{split}
\hat{S} = \operatorname{Tokenizer}(\hat{U}; \varphi), \qquad
S = \operatorname{Tokenizer}(U; \varphi),
\end{split}
\end{equation}
where $\varphi$ are parameters of the subword tokenizer, which is trained on a large speech corpus.
Then the SpeechBLEU score is computed on $\hat{S}$ and $S$.

We used SentencePiece for subword tokenization using BPE and unigram language model tokenization.
Table~\ref{tab:eval-subword} lists the results.
For SpeechBLEU, we used cases with $G=2$, excluding token repetition.
For SpeechTokenDistance (Jaro-Winkler), token repetition was not excluded.
We also varied the vocabulary size of subword tokenization from 256 to 4096.
The results indicate that subword tokenization does not improve the correlation.
Furthermore, the performance of both SpeechBLEU and SpeechTokenDistance decreased as the vocabulary size increased.
This suggests that capturing the fine-grained structure of speech token sequences is important for automatic evaluation tasks.
In the case of SpeechBLEU, the increased sparsity of subword tokens can make $n$-gram matches rarer, reducing the reliability of BLEU scores.

\section{Evaluations with unaligned reference speech}\label{sec:appendix-unaligned}

\begin{figure}
    \centering
    \includegraphics[width=0.60\linewidth, clip]{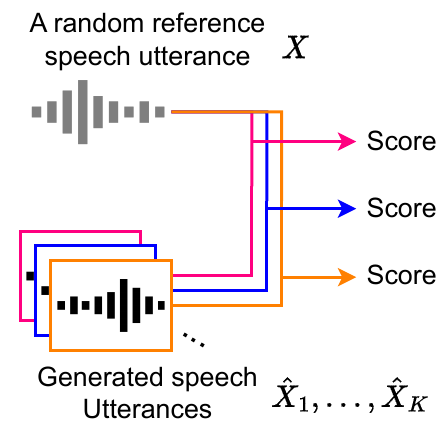}
    \caption{Illustration of evaluations with unaliend reference speech.}
    \label{fig:unaligned}
\end{figure}

\begin{table}[tb]
\centering
\caption{Results with unaligned references.}
\label{tab:eval-unaligned}
\begin{tabular}{lcccc}
\toprule
\multicolumn{1}{l|}{\multirow{2}{*}{}} & \multicolumn{2}{c|}{Utterance-leven}                 & \multicolumn{2}{c}{System-level} \\
\multicolumn{1}{l|}{}                  & LCC            & \multicolumn{1}{c|}{SRCC}           & LCC             & SRCC           \\ \midrule
\multicolumn{5}{l}{\textit{Traditional metrics (Unaligned reference)}}                                                           \\ \midrule
\multicolumn{1}{l|}{MCD}               & 0.0066         & \multicolumn{1}{c|}{0.026}          & 0.082           & 0.122          \\ \midrule
\multicolumn{5}{l}{\textit{Unsupervised, without reference}}                                                                     \\ \midrule
\multicolumn{1}{l|}{SpeechLMScore~\cite{maiti2022speechlmscore}}     & 0.164          & \multicolumn{1}{c|}{0.127}          & 0.268           & 0.246          \\ \midrule
\multicolumn{5}{l}{\textit{Supervised, without reference}}                                                                       \\ \midrule
\multicolumn{1}{l|}{UTMOS~\cite{saeki2022utmos}}             & \textbf{0.363} & \multicolumn{1}{c|}{0.340}          & 0.537           & 0.575          \\ \midrule
\multicolumn{5}{l}{\textit{Proposed SpeechBERTScore (Unaligned reference)}}                                                      \\ \midrule
\multicolumn{1}{l|}{\texttt{wav2vec2-base}}     & 0.101          & \multicolumn{1}{c|}{0.109}          & 0.149           & 0.207          \\
\multicolumn{1}{l|}{\texttt{hubert-base}}       & 0.075          & \multicolumn{1}{c|}{0.083}          & 0.074           & 0.078          \\
\multicolumn{1}{l|}{\texttt{wavlm-base}}        & 0.075          & \multicolumn{1}{c|}{0.087}          & 0.074           & 0.105          \\
\multicolumn{1}{l|}{\texttt{wavlm-base+}}       & 0.097          & \multicolumn{1}{c|}{0.108}          & 0.100           & 0.110          \\ \midrule
\multicolumn{1}{l|}{\texttt{wav2vec2-large}}    & 0.119          & \multicolumn{1}{c|}{0.201}          & 0.277           & 0.376          \\
\multicolumn{1}{l|}{\texttt{hubert-large}}      & 0.215          & \multicolumn{1}{c|}{0.306}          & 0.473           & 0.535          \\
\multicolumn{1}{l|}{\texttt{wavlm-large}}       & 0.268          & \multicolumn{1}{c|}{\textbf{0.360}} & \textbf{0.543}  & \textbf{0.606} \\ \bottomrule
\end{tabular}
\end{table}

In \S~\ref{sec:eval}, we dealt with cases where reference speech $X$ and generated speech $\hat{X}$ are aligned, meaning they have the same spoken content.
This section evaluates cases where $X$ and $\hat{X}$ are unaligned, with different spoken content.
This is an automatic evaluation framework capable of assessing the naturalness of a group of synthesized speech samples from a given speaker, using just one sample of the speaker's natural voice.
Such automatic evaluation is crucial in assessing zero-shot TTS~\cite{wang2023neural} and speech continuation~\cite{borsos2023audiolm}.

As shown in Fig.~\ref{fig:unaligned}, we randomly selected a natural speech utterance as the reference speech $X$.
Let $\hat{X}_{1}, \cdots \hat{X}_{k}, \cdots \hat{X}_{K}$ denote all the synthetic speech samples, where $K$ is the number of synthetic speech samples.
As we used the SOMOS datasets described in \S~\ref{sec:eval-setting-dataset}, $K=1000$ was satisfied.
Table~\ref{tab:eval-unaligned} lists the results.
Note that the results from SpeechLMScore and UTMOS are the same as those in \S~\ref{sec:eval-main} as they do not use reference speech.
MCD showed a very low correlation in cases where reference and generated speech are unaligned.
For the proposed SpeechBERTScore, smaller models like \texttt{wav2vec2-base} showed lower correlation. Interestingly, larger models like \texttt{wavlm-large} significantly improved correlation.
In utterance-level LCC, UTMOS performed best, but unaligned SpeechBERTScore with \texttt{wavlm-large} surpassed UTMOS in other metrics, suggesting the feasibility of using SpeechBERTScore for evaluations with unaligned reference speech.
Further analysis of this unaligned speech evaluation will be the future work.

\end{document}